# Pauli blocking of atomic spontaneous decay

Christian Sanner, Lindsay Sonderhouse, Ross B. Hutson, Lingfeng Yan,

William R. Milner, and Jun Ye

JILA, National Institute of Standards and Technology, and University of Colorado,

Boulder, CO 80309, USA

**Spontaneous decay of an excited atomic state is a fundamental process that originates from the interaction between matter and vacuum modes of the electromagnetic field. The rate of decay can thus be engineered by modifying the density of final states of the joint atom-photon system. Imposing suitable boundary conditions on the electromagnetic field has been shown to alter the density of vacuum modes near the atomic transition, resulting in modified atomic decay rates. Here we report the first experimental demonstration of suppression of atomic radiative decay by reducing the density of available energy-momentum modes of the atomic motion when it is embedded inside a Fermi sea.**

Spontaneous radiative decay of an excited quantum system is one of the most fundamental processes in nature. The phenomenon is ubiquitous: it makes fireflies glow, underlies the radiative recombination of electrons and holes in light-emitting diodes, and is responsible for the gamma decay of nuclear isomers. The quantum theory of spontaneous emission (*1*) interprets the relaxation process of an excited atom as an interplay between the quantized electromagnetic field and atomic states. Driven by vacuum fluctuations (*2*), a single atomic excitation can decay by emitting a photon into a multitude of electromagnetic field modes so that the process is irreversible. Manipulation of the density of vacuum modes through the use of an electromagnetic resonator modifies the emission and scattering of light. In particular, a resonant cavity can increase the spontaneous emission rate while a mistuned resonator can inhibit spontaneous decay (*3–5*). The natural lifetime of an excited atomic state thus depends on the atom's electromagnetic environment. This Purcell effect is now widely used in nanostructured devices (*6*).

Since the spontaneous decay depends on the final density of states for the joint atom-photon system, there must be an analogous effect where the mode structure of final atomic motional states can affect the spontaneous decay of the excited atom. More than 30 years ago it was suggested (*7*) that constraints imposed by quantum statistics could modify spontaneous emission. Fermi statistics requires the total wavefunction of a fermionic system to be antisymmetric, giving rise to the Pauli exclusion principle that forbids indistinguishable fermions from occupying the same internal and external quantum states. Accordingly, if a sufficient number of ground state fermionic atoms occupy all available external motional states into which an internally excited fermionic atom has to decay, this decay process will be blocked. The prospect to quantum engineer the natural lifetime of an excited atomic state by embedding it inside a Fermi sea has triggered many theoretical studies and proposals (*8–15*), but up to now this fundamental quantum effect has not been observed, complicated by atomic properties and competing collective radiative behavior.

Here, we report the first direct observation of Pauli suppression of spontaneous emission using a quantum degenerate Fermi gas of strontium atoms. We confirm that the suppression becomes stronger as degeneracy is increased and when the Fermi energy approaches the photon recoil energy. By angularly resolving the incoherent photon scattering rate, we measure up to a factor-of-two reduction in comparison to the natural value determined from a thermal ensemble. Averaged over all emission directions this corresponds to a ~10% increased natural lifetime of the excited state. This striking manifestation of Fermi statistics connects for the first time the fundamental radiative property of atoms to their motional degrees of freedom subject to quantum statistics. The consequences of Pauli blocking of atomic motion have been demonstrated earlier, including the suppression of collisions (*16*), the direct observation of Fermi pressure (*17*), the onset of Hanbury Brown-Twiss anticorrelations (*18*, *19*), local antibunching (*20–22*), the suppression of chemical reactions between molecules (*23*), and the formation of Pauli crystals (*24*).

Fig. 1 illustrates the key concept of the experiment and introduces the relevant energy scales (*25*). An ensemble of *N* harmonically confined identical fermions of mass *m* forms a quantum degenerate Fermi sea with close to unity occupation of the oscillator states if the thermal energy $k_B T$ is small compared to the Fermi energy $E_F = (6N)^{1/3} \hbar\omega$. The three-dimensional confinement is characterized by the mean trap frequency $\omega = (\omega_x \omega_y \omega_z)^{1/3}$, and $k_B$ and $\hbar$ denote the Boltzmann and reduced Planck constants, respectively. The Fermi temperature $T_F$ and Fermi wavevector $k_F$ are defined via $E_F = (\hbar k_F)^2 / (2m) = k_B T_F$. If an atom inside the Fermi sea absorbs a photon carrying momentum $\hbar \mathbf{k}_{abs}$, it gains a corresponding recoil energy $E_R = (\hbar k_{abs})^2 / 2m$. Here, we consider the case of weak confinement with $\hbar\omega << E_R$. Upon spontaneous photon emission the atom experiences a randomly directed second momentum kick $\hbar \mathbf{k}_{emi}$ where $|\mathbf{k}_{emi}| = |\mathbf{k}_{abs}| = k_R$, resulting in a total momentum transfer $\hbar \mathbf{k} = \hbar \mathbf{k}_{abs} + \hbar \mathbf{k}_{emi}$. If, however, the corresponding motional state is already occupied by another atom within the Fermi sea, this decay cannot happen, and light scattering will be suppressed. The relative temperature $T/T_F$ and the wavevector ratio $k/k_F$, where $\hbar k_F = (2mE_F)^{1/2}$ is the momentum space radius of the Fermi sea, determine the density of available final momentum states and hence the degree of blockade. Following Fermi's golden rule one finds using a local density approach (*11, 14*) a relative scattering rate of

$$S(\mathbf{k}) = \frac{\int d^3\mathbf{p}\, d^3\mathbf{q}\, n_i(\mathbf{p},\mathbf{q})[1 - n_f(\mathbf{p},\mathbf{q})]}{\int d^3\mathbf{p}\, d^3\mathbf{q}\, n_i(\mathbf{p},\mathbf{q})}.$$

Here the integrals cover the six-dimensional phase space spanned by three momentum dimensions $\mathbf{p}$ and three real space dimensions $\mathbf{q}$. The initial and final state phase space cell occupations are given by $n_i = n_{FD}(\mathbf{p}, \mathbf{q})$ and $n_f = n_{FD}(\mathbf{p}+\hbar\mathbf{k}, \mathbf{q})$, where $n_{FD}$ is the Fermi-Dirac distribution for a harmonically trapped gas, that is,

$$n_{FD}(\mathbf{p},\mathbf{q}) = \frac{1}{1 + \xi^{-1}\exp[\{\sum_i p_i^2/(2m) + \sum_i m\omega_i^2 q_i^2/2\}/k_B T]}.$$

The index $i$ runs over all three dimensions and $\xi$ is the fugacity related to $T/T_F$ via $1 / \text{Li}_3(-\xi) = -6\,(T/T_F)^3$, where $\text{Li}_3$ is the trilogarithm function. The expression for $S(\mathbf{k})$ counts all available final momentum states for a given momentum transfer $\hbar\mathbf{k}$ and averages over all initial states within the Fermi sea.

The above analysis suggests two pathways to observe pronounced Pauli suppression of radiative decay. One can either prepare a Fermi gas with $E_F \gg E_R$ so that a significant decay blockade is obtained for all emission directions, that is, for any momentum transfer up to the maximum $2\hbar k_R$ (homogeneously colored right sphere in Fig. 1C). Or one can relax this requirement and selectively observe only scattering events with a small momentum transfer so that $E_F \sim E_R$ is sufficient. The second approach, which we take here (gradient-colored left sphere in Fig. 1C), is straightforwardly realized in a small-angle light scattering configuration where a small number of atoms within the Fermi sea are optically excited and a subset of the spontaneously emitted photons is collected under a shallow angle $\alpha$ with respect to the excitation beam corresponding to a momentum transfer of $\hbar k = 2\hbar k_R \sin(\alpha/2)$.

A multitude of effects besides quantum statistics can influence radiation dynamics in a dense ensemble of emitters. Coherence, either externally imprinted or spontaneously established, can lead to super- and subradiant collective states that correspondingly exhibit super- and subnatural radiative lifetimes. Dicke superradiance (*26*, *27*), radiation trapping (*28*), multiple scattering (*29*), and other forms of coherent or incoherent collective scattering (*30*) all critically depend on the integrated optical density $OD = \int \sigma n \, dl$ of the atomic gas. This attenuation parameter, derived from the single-atom scattering cross section $\sigma$ and density $n$, where $l$ is measured along the direction of the incoming light, defines a parameter region where the gas is optically thin ($OD \ll 1$) and single-particle scattering dominates over collective effects. Given that on resonance, $\sigma$ is on the order of the squared optical wavelength $\lambda^2 = (2\pi/k_R)^2$, and $k_F$ is tied to the 3D peak density via $n = k_F^3/6\pi^2$, it is not possible to satisfy the Pauli blocking criterion $k_F \sim k_R$ without violating the small *OD* requirement. Indeed, typical optical densities encountered in atomic Fermi gas experiments easily exceed 100. In this optically thick regime multiple photon scattering strongly affects the light propagation inside the sample, as evidenced with the resonant fluorescence image displayed in Fig. 2B. Using off-resonant light reduces the effective scattering cross section and renders the atom cloud weakly absorbing for the incoming probe light at sufficient detuning. This, combined with differential observation strategies and avoiding light detection in the forward direction, minimizes the influence of collective scattering dynamics. Furthermore, unlike Pauli blocking, these competing effects are not dependent on quantum statistics and can therefore be observed using a thermal gas (*31*), instead of a quantum degenerate gas, to provide a baseline.

Our experiment starts with the preparation of a $^{87}$Sr Fermi gas as described in reference (*32*). The $^1S_0$ $F = 9/2$ ground state is split into ten magnetic spin states $m_F = -9/2, ..., 9/2$. Here $F$ is the total angular momentum of the nuclear spin. This fully thermalized 10-component sample (*33*) contains 18 000 atoms per spin state confined in a crossed optical dipole trap with maximum radial trap frequencies of $\omega_x = \omega_y = 2\pi \times 120$ Hz and an axial confinement with $\omega_z = 2\pi \times 506$ Hz. This leads to a Fermi energy of $E_F = 440$ nK for each of the ten Fermi seas and we reach temperatures down to 0.1 $T_F$. Under suitable conditions, any optical excitation returning to the ground state should experience a decay blockade. We perform specific measurements on the $^1S_0$ - $^1P_1$ transition at 461 nm with a natural linewidth (*34*) of $\Gamma = 2\pi \times 30.4$ MHz and a recoil energy of $E_R = 520$ nK. Instead of resolving decay dynamics on the scale of nanoseconds ($1/\Gamma = 5.2$ ns), we devise a continuous weak-drive scheme where the Fermi gas is exposed for 1 μs to a 1.2 GHz detuned drive beam that causes on average <10% of the atoms to undergo an excitation cycle.

For the given drive beam and atom cloud parameters the sample is optically thin with an effective OD of 0.02, corresponding to a forward transmission of 98%. As illustrated in Fig. 2, two detectors collect scattered photons simultaneously under off-axis angles of $\alpha_1 = 24°$ and $\alpha_2 = 72°$. These operational parameters are chosen to simplify the interpretation of our measurements. First, as discussed above, the detuning from resonance by 40$\Gamma$ eliminates multiple scattering dynamics and furthermore avoids refractive lensing contributions (*35*) far off-axis. Second, the low excitation rate keeps the Fermi sea intact throughout the probe pulse exposure. Third, operating in the weak-drive limit ensures that there is no inelastic scattering contribution beyond the recoil-induced energy shift, i.e., Mollow triple-peaked fluorescence spectra and other strong-drive effects are negligible (*36*).

To systematically explore how quantum degeneracy affects light scattering, we start with a deeply degenerate gas of $T/T_F = 0.1$ and gradually heat it up to $T/T_F = 0.7$ through parametric confinement modulation while keeping the atom number and Fermi energy constant. Under these conditions incoherently scattered photons excited by the weak probe pulse are counted. To satisfy the requirement to minimally disturb the Fermi gas by the probe, the number of collected photons along the two detection axes is correspondingly low. Even for the axis with a high NA = 0.23 objective lens fewer than 200 photons are collected at full quantum efficiency. By operating the gated CCD detectors in a hardware binning mode that maps all detected photons into a 3x3 superpixel array, we spatially and temporally isolate the signal from background contributions and maintain low readout noise. The measured photon counts are compatible with ab-initio estimates within 30% based on the reported drive parameters and detection efficiencies. All relevant thermodynamic parameters are independently assessed through measurements on expanded gas clouds after time of flight.

The results of Pauli-suppressed scattering are shown in Fig. 3A. Under the shallow off-axis angle of 24°, where $k/k_F = 0.45$ for our Fermi sea, we find a strong dependence of the photon counts on $T/T_F$. To properly normalize the detector signal, i.e., to convert the registered photon counts to a suppression ratio without introducing an arbitrary scaling factor, it is necessary to prepare an equivalent reference sample that is not Pauli-blocked. Because the weak optical confinement does not maintain a constant atom number for $T/T_F > 1$, we devise an alternative method to eliminate Pauli blocking: At 10 μs before applying the actual probe pulse we expose the Fermi gas to a 5 μs long pre-pulse that destroys the Fermi sea by randomly exciting atoms to momentum states beyond $k_F$. Consequently, light scattering is no longer Pauli suppressed during the subsequent probe pulse. All blue round data points in Fig. 3A/B are normalized to a common reference photon count obtained from a single pre-pulse exposed sample as detailed in the Supplementary Materials (*37*). Numerically integrating the expression for $S(\boldsymbol{k})$ at all probed temperatures reveals good agreement between experiment and theory. The simultaneously acquired measurements at an observation angle of 72° do not show a pronounced suppression and only exhibit a weak temperature dependence, as expected for $k/k_F = 1.27$ since most final momentum states lie outside of the Fermi sea.

To further verify that Pauli blocking is the mechanism responsible for the observed scattering behavior, we study the dependence on $k_F/k_R$ by varying the confinement while keeping atom number and $T/T_F = 0.13$ constant, as displayed in Fig. 3B. Even at the shallowest confinement with $k_F/k_R = 0.57$ ($E_F/E_R = 0.32$), the momentum transfer along the 24° axis amounts to only $k/k_F = 0.74$ so that we still observe substantial suppression of light scattering. This is in contrast to the 72° case, where $k/k_F = 2.07$ at $k_F/k_R = 0.57$. If $k/k_F > 2$ Pauli blocking is negligible at any temperature so that we normalize all photon counts acquired under 72° (red data points in Fig. 3A/B) with respect to this reference point (solid red circle in Fig. 3B).

The additional blue square data points displayed in Fig. 3 were normalized by applying a pre-pulse separately for each data point and are in good agreement with the common-mode normalized measurements. In particular, with a 10 µs wait time before the probe pulse, the atomic density remains essentially the same with or without the pre-pulse. This ensures common-mode cancelation of collective scattering dynamics. In the Supplementary Materials (*37*) we present light scattering measurements with variable pre-pulse durations for a deeply degenerate ($T/T_F$ = 0.11) and heated ($T/T_F$ = 0.58) Fermi gas. This data confirms the scattering-induced destruction of the Fermi sea over the course of the pre-pulse application.

Spatially resolving the origin of the scattered photons within the Fermi gas, in addition to counting the total number of scattered photons along a given direction, provides a picturesque revelation of the Pauli blocking mechanism. For this purpose, we modify the 24° high-NA axis to deliver magnified images of the atom cloud and employ a non-binning CMOS camera with a low readout noise of 2 photoelectrons. The cylindrically symmetric atomic cloud at $T/T_F$ = 0.12 with a diameter of about 20 μm is projected along the *z*-axis into a 2D image with 0.9 μm wide pixels. Because a single pixel collects on average less than 1 photon, it is necessary to average hundreds of frames in order to derive a finely resolved scattering profile. Furthermore, we radially average the mean image to obtain the radial profile (blue data points) shown in Fig. 4. Residual optomechanical drifts in the optical setup cause small displacements of the center of mass position of the cloud during the frame averaging period. This, together with the finite resolution of the imaging system, leads to an effective *$1/e^2$* pixel blurring on the order of 3 μm. To compare the observed scattering profile to theory predictions without introducing a free scaling parameter, we independently acquire in-situ density profiles of the Fermi gas through the same imaging setup via high intensity fluorescent imaging (*38*,

*39*). Using the frame-wide summed total photon counts and the measured global suppression ratio at $T/T_F = 0.12$ (Fig. 3A), we then properly scale the radial density profile to reflect the scattering profile expected without Pauli blockade for the weak probe beam (purple curve). Momentum space integration plus one-dimensional line-of-sight integration of $S(\boldsymbol{k})$ yields a radially resolved suppression ratio for us to determine the expected scattering profile with Pauli blockade (blue curve). Except at the center of the cloud, where radial averaging does not significantly improve the signal-to-noise ratio, we find good agreement between calculated and measured profiles. Towards the outer rim of the cloud the local Fermi energy drops so that light scattering is not suppressed anymore; the local suppression ratio will approach the thermal gas limit of 1. This happens, as seen with the blue theory curve, only in the outermost region where the density is so low that the signal-to-noise level is insufficient to reliably determine a suppression ratio.

In conclusion, we have reported a clear demonstration that Fermi statistics leads to strongly modified light scattering in a quantum degenerate system. The presence of a Fermi sea alters the final atomic motional mode spectrum and enables a direct observation of Pauli blockade of spontaneous decay. Interpreted from a many-body system perspective, this experiment probes the structure factor (*40*) of a quantum degenerate Fermi gas. Defined as the Fourier transform of the spatial density-density correlation, the static structure factor $S(\boldsymbol{k})$ characterizes the linear response of a system to a perturbation with wavevector $\boldsymbol{k}$. Accordingly, the suppression of light scattering and the suppression of density fluctuations (*21*) in a Fermi gas are two interrelated manifestations of the same fundamental many-body physics.

It will be exciting to exploit the capability of altering a fundamental decoherence mechanism for opening new avenues for quantum engineered atom-light interfaces. In particular, custom designed Fermi reservoirs can protect optical qubits at local nodes while facilitating cavity-free directional photon emission for efficient network connectivity. In the context of optical atomic clocks this work could enable spectroscopic interrogation times exceeding the natural lifetime of the excited clock state and investigation of novel radiative properties of atoms.

We thank Peter Zoller, Ana Maria Rey, James Thompson and Murray Holland for stimulating discussions and careful reading of the manuscript. Funding for this work is provided by DARPA, NSF QLCI OMA–2016244, NSF Phys-1734006, and NIST. C.S. thanks the Humboldt Foundation for support.

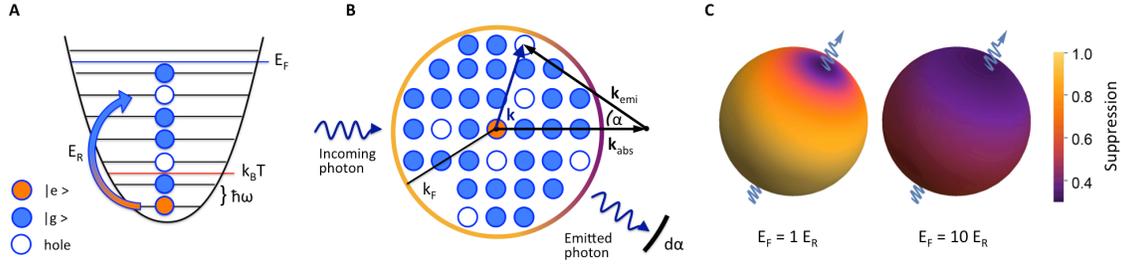

**Fig. 1. Spontaneous decay of an atom embedded inside a Fermi sea.**

(**A**) Indistinguishable fermions obey the Pauli exclusion principle. If the thermal energy $k_B T$ is sufficiently low they fill almost all available harmonic oscillator states up to the Fermi energy $E_F$ with near-unity occupation. An excited atom (orange) acquires a recoil energy $E_R$ when returning to its internal ground state (blue). (**B**) In momentum space, the atoms form a Fermi sea occupying most states up to the Fermi momentum $\hbar k_F$. Spontaneous decay of an excited atom with emission along $\alpha$ and total momentum transfer $\hbar \mathbf{k}$ can happen only if the final momentum state is not occupied by another ground state fermion. A detector covering a solid angle $d\alpha$ registers the emitted photon. (**C**) Pauli blocking leads to a characteristic angular distribution of scattered photons in the deeply degenerate regime (here $T/T_F = 0.1$). For $E_F \sim E_R$ (left sphere) scattering is preferentially suppressed in a small cone around the forward direction, while $E_F \gg E_R$ (right sphere) causes strong suppression for all scattering angles $\alpha$. The suppression factor specifies the scattering rate relative to a non-Pauli-blocked sample.

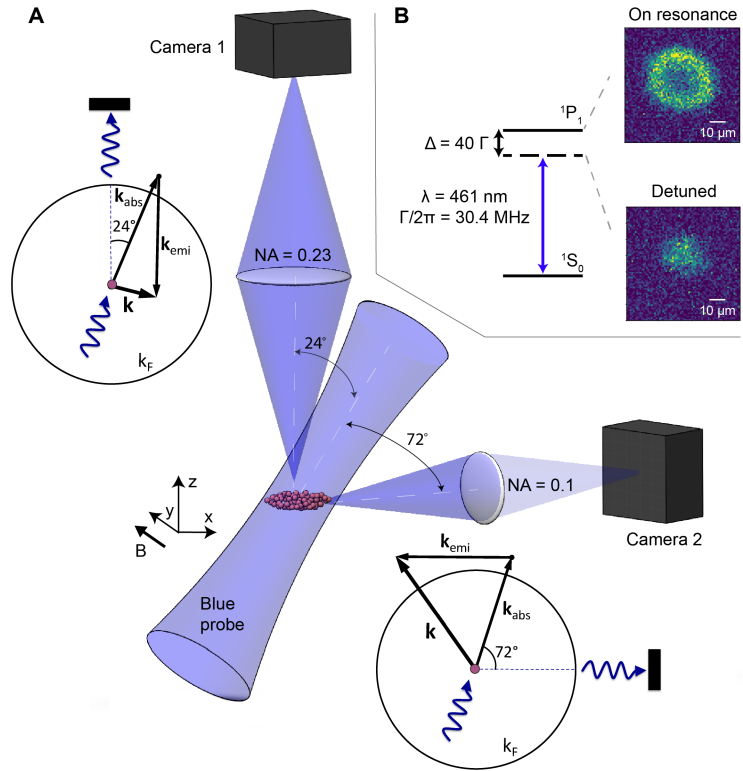

**Fig. 2. Experimental setup.**

(**A**) Off-resonant probe light excites $^{87}$Sr atoms inside a Fermi sea. Spontaneously reemitted photons are collected simultaneously along two imaging axes under angles of 24° and 72°, with their numerical apertures (NA) shown respectively. Small scattering angles correspond to a small momentum transfer with $k/k_F < 1$, whereas the transversal observation detects photons from scattering events with $k/k_F > 1$. The circularly polarized probe beam has an intensity of 5 $I_{sat}$ where the resonant saturation intensity is $I_{sat} = 41$ mW/cm$^2$. (**B**) On resonance the atomic cloud is optically thick for the probe beam and the image formed on Camera 1 displays a hole in the cloud center due to multiple scattering. At a detuning of $\Delta = 40\Gamma$ the atom cloud is optically thin and the corresponding image resembles the atomic density distribution. The detuned frequency is used in the Pauli blocking experiment. A magnetic bias field of 3 Gauss applied in the horizontal plane along the *y*-direction defines the atomic quantization axis.

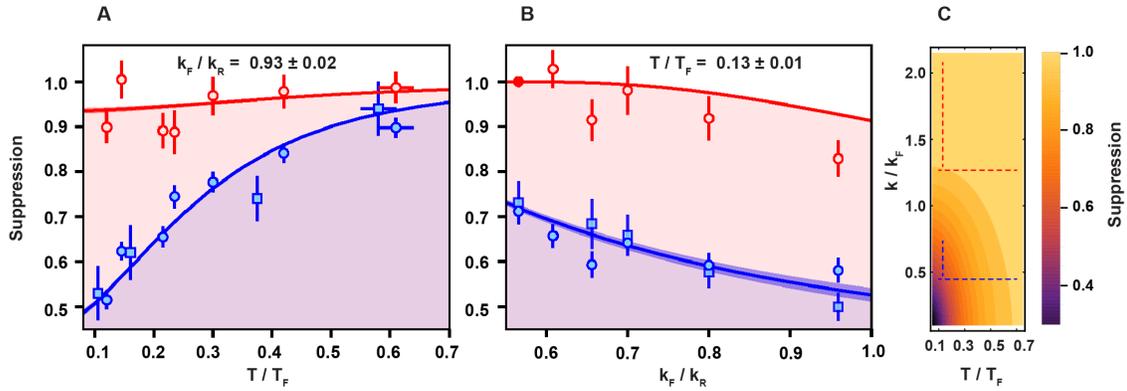

**Fig. 3. Suppression of light scattering in a $^{87}$Sr Fermi gas over a range of temperatures and Fermi momentums.**

All measurements are performed with a 10-component Fermi gas containing 18 000 atoms per spin state. The scattering behavior is distinctly different for the two observation angles of 24° (blue circles and squares) and 72° (red circles). Raw photon counts are normalized with respect to measurements on non-Pauli-blocked reference samples (see main text). Each circle data point is derived from 150 iterations of the experiment, while each square point results from 50 experimental runs. Solid theory curves are calculated with no free parameters. The widths of the theory lines reflect the experimental uncertainties of Fermi energy and temperature. The error bars are purely statistical and indicate one-standard-deviation confidence intervals. (**A**) At a constant Fermi wavevector of $k_F/k_R = 0.93$ ($E_F/E_R = 0.86$), the atom ensemble's scattering cross section decreases as the gas approaches deep quantum degeneracy. The suppression observed under 24° is pronounced and reaches 50% at $T/T_F = 0.13$. In contrast, under 72°, the suppression is negligible. (**B**) At constant $T/T_F = 0.13$, $k_F$ is varied by adiabatically changing the confinement. A larger $k_F$ results in a stronger suppression. (**C**) The data reported in **A** and **B** are measured along 4 trajectories (dotted lines) through the parameter space spanned by $k/k_F$ and $T/T_F$. Depending on the scattering angle $k$ varies between 0 and $2k_R$. Light collected under an off-axis angle of 24° corresponds to a momentum transfer $\hbar k < \hbar k_F$ for the given Fermi gas, leading to substantial reduction of the density of available final states. On the contrary, for the 72° collection angle, the corresponding momentum transfer $\hbar k > \hbar k_F$. Thus most final states are not blocked and scattering is not suppressed.

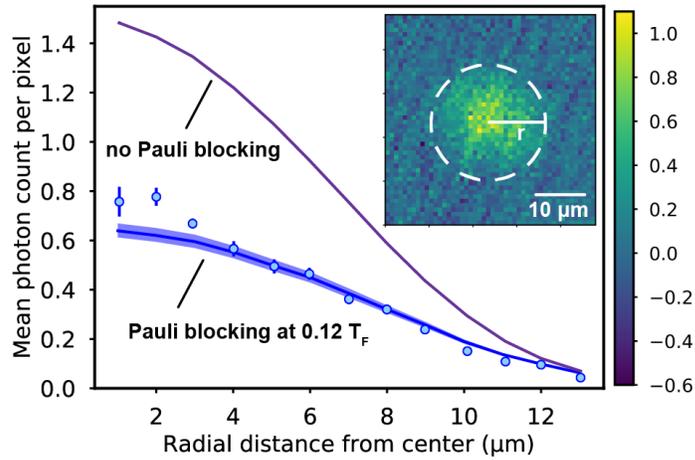

**Fig. 4. Spatially resolved light scattering from a trapped Fermi gas at $T/T_F = 0.12$.** Radially averaging the spatially resolved mean signal (inset) from 1100 individual images obtained along the z-axis yields a radial light scattering profile (blue data points). In-situ column density images, separately obtained using a high intensity fluorescent imaging technique, are used to predict the scattering signal for a non-degenerate gas (purple curve). The spatial profile of light scattering calculated for the $T/T_F = 0.12$ ensemble (blue curve) agrees well with the measured data.

# Pauli blocking of atomic spontaneous decay

## Supplementary Materials

Christian Sanner, Lindsay Sonderhouse, Ross B. Hutson, Lingfeng Yan,

William R. Milner, and Jun Ye

JILA, National Institute of Standards and Technology, and University of Colorado,

Boulder, CO 80309, USA

The suppression factors reported in Fig. 3 as red and blue circles were obtained by normalizing the measured photon counts against a common reference count value. Residual total atom number and probe pulse intensity fluctuations were below 10% over the course of the data acquisition. For the 72° detection axis (red circles) the measurement at $k_F/k_R = 0.57$ (red solid circle), which is outside the Pauli-blocking regime with $S > 0.99$, directly provided the reference count. The photon counts detected along the 24° axis were normalized against a reference count measured by exposing the Fermi gas to a 1.2 GHz detuned pre-pulse to create holes in the Fermi sea before applying the actual probe pulse. Precise detection gating with an interline transfer CCD (Camera 1) and a short dark time of 10 μs between the two pulses avoided contamination of the probe signal with pre-pulse fluorescence while keeping the atomic density distribution unaffected. The pre-pulse intensity was calibrated such that each atom scattered on average more than one photon during a 5 μs long pulse. To directly verify that this exposure destroyed the Fermi sea and eliminated the Pauli blockade, we systematically varied the pre-pulse duration and observed for two samples at $T/T_F = 0.11$ and $T/T_F = 0.58$ how the probe signal saturated at the non-blocking reference level for longer pre-pulses (Fig. S1). A pre-pulse duration of 5 μs was chosen to obtain the reference count along the 24° detection axis.

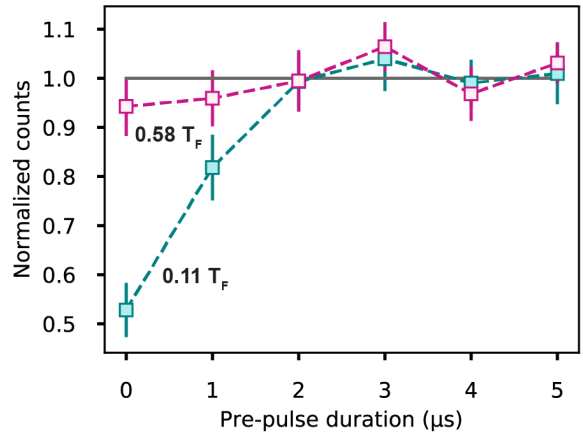

**Fig. S1. Photon scattering from a Fermi gas after exposure to a pre-pulse of variable duration.**

For the chosen pre-pulse detuning and intensity, the Pauli blockade is destroyed after exposure for a few μs. The scattering signal from the deeply degenerate sample (green squares) increases by almost a factor of 2 while the barely degenerate sample (purple squares) shows only minimal increase, as expected for a Fermi sea with $k_F/k_R = 0.93$ under an observation angle of 24°. Data in the plot is normalized relative to the mean counts detected for 4 and 5 μs pre-pulse durations.